# A PRODUCT ORIENTED MODELLING CONCEPT
## Holons for systems synchronisation and interoperability

Salah Baïna, Hervé Panetto
*CRAN-UMR 7039, Nancy-University, CNRS*
*F 54506 Vandoeuvre les Nancy, France,*
*salah.baina@cran.uhp-nancy.fr, herve.panetto@cran.uhp-nancy.fr,*
*gerard.morel@cran.uhp-nancy.fr*

Khalid Benali
*LORIA-UMR 7503), Nancy-University, CNRS, INRIA*
*F 54506 Vandoeuvre les Nancy, France,*
*benali@loria.fr*

Abstract: Nowadays, enterprises are confronted to growing needs for traceability, product genealogy and product life cycle management. To meet those needs, the enterprise and applications in the enterprise environment have to manage flows of information that relate to flows of material and that are managed in shop floor level. Nevertheless, throughout product lifecycle coordination needs to be established between reality in the physical world (physical view) and the virtual world handled by manufacturing information systems (informational view). This paper presents the "Holon" modelling concept as a means for the synchronisation of both physical view and informational views. Afterwards, we show how the concept of holon can play a major role in ensuring interoperability in the enterprise context.

## 1. INTRODUCTION

Enterprise application integration (EAI) and the opening of information systems towards integrated access have been the main motivation for the interest around systems interoperability. Integration aspect and information sharing in the enterprise lead to an organisation of the hierarchy of enterprises applications where interoperability is a key issue (see Fig. 1).

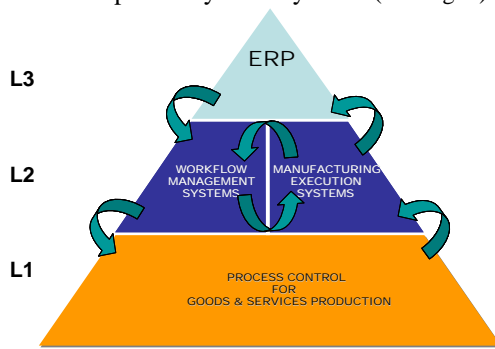

Fig. 1 Manufacturing enterprises common structure

This hierarchy defines the three main levels in manufacturing enterprises:

L1: Process control level contains all processes that perform routing and physical transformations on the produced goods and services;

L2: The Execution level performs the processes that manage decision flows (e.g.: Workflow systems) and production flows (e.g.: MES[1], SCE[2]);

L3: The management system level is responsible for the management of processes that handle all different informational aspects related to the enterprise (e.g.: APS[3], ERP[4] or CRM[5] systems).

To meet traceability, product genealogy and product life cycle management needs, nowadays an enterprise has to manage flows of information that relate to flows of material and that are managed in shop floor level. We assume that the enterprise is composed of two separated worlds (see Fig. 2):

*(i)* On one hand, a world in which the product is mainly seen as a physical object, this world is

---
[1] Manufacturing Execution System
[2] Supply Chain Execution
[3] Advanced Planning System
[4] Enterprise Resource Planning
[5] Client Relationship Management

called the manufacturing world. It handles systems that are tightly related to the shop-floor level,

(ii) On the other hand, a world where the product is seen as a service released in the market. This world is called the business world.

In order to achieve the main objective of the enterprise, "the product" to be specific, the business universe and the manufacturing universe need to exchange information and to synchronise their knowledge concerning the product (good and service). It is assumed that the product (good/service) can play the role of the gateway between both universes, since it represents a common entity between those worlds.

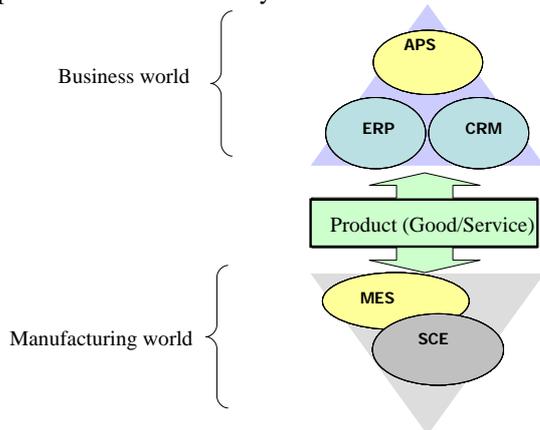

Fig. 2. Product centric approach

In this paper, we define a holon based approach in order to synchronise views in the business world and in the physical manufacturing world using the holon concept. The paper continues by presenting the usability of the concept of holon in ensuring interoperability enterprise context. Section 2 presents the bases of our holonic process modelling concepts (Morel, *et al.*, 2003) that use the product as a centric entity in process models. Section 3 of the paper gives a brief introduction to interoperability in the enterprise and explains how holons can be used as a means for enterprise applications interoperability. In Section 4, an implementation of the holon is proposed. Section 5 gives conclusions and perspectives for this work.

## 2. A MODELLING CONCEPT FOR PRODUCT REPRESENTATION

In this section, we introduce the holon as a modelling concept. Afterwards, we will show how this concept can be exploited in order to facilitate taking into account interoperability concerns in modelling phase. Existing solutions for interoperability in enterprise environment focus mainly on enterprise processes interoperability and interconnection. Throughout product lifecycle, coordination needs to be established between the reality in the physical world where the product evolves as a physical object and the "electronic" world handled by manufacturing information systems where the virtual image of the product evolves as an informational object. Our work aims to provide a product centric approach for enabling interoperability between information systems in the manufacturing environment in order to establish the coherence between the physical products and their informational representations. To take into account this duality (physical things/ informational things), we propose an adaptation of the concept of holon (Koestler, 1967) to this specific problem.

### 2.1 The Holon Concept in Manufacturing Process Modelling

The word Holon is a combination of the Greek word *holos*, meaning whole, and the suffix *on* meaning particle or part. A holon is an identifiable part of a system that has a unique identity, yet is made up of sub-ordinate parts and in turn is part of a larger whole. A Holon has two main features, autonomy and cooperation. Several adaptations of the holon concept have been proposed in several fields. In the manufacturing context, a Holonic Manufacturing System (HMS) is an autonomous and co-operative building block of a system for transforming, transporting, storing and/or validating information and physical objects (Mc Farlane and Bussmann, 2000; Seidel and Mey, 1994). In this paper, we adapt the *holon* concept definition to solve the problem of synchronisation between physical views and informational views of the same objects. We define the holon then as an aggregation of an information part and a physical part.

### 2.2 The Holon definition:

In Holonic Process Modelling (Morel, *et al.*, 2003; Baïna, *et al.,* 2005; Valckenaers, 2001).), holons are used to represent products; the physical part of the holon represents the material part (also called physical view) of the product and the informational part of the holon represents the informational part

(informational view) of the product. Characteristics of holon are distinguished into two categories;
- Attributes describing the current state of the holon. The state of a holon contains three kinds of attributes: space attributes, shape attributes, and time attributes (Panetto and Pétin, 2005);
- Properties related to the holon but which do not correspond to any of the three types of properties; space, shape or time.

Holons can be classified into two categories; *(i)* elementary holons and *(ii)* composite holons:
*(i)* Elementary holons are the combination of a single informational part and a single physical part.
*(ii)* Composite holons are the result of the processing and treatment of one or more other holons, this processing can be the aggregation of a set of holons (composite or elementary) in order to compose a new holon or a transformation of one composite holon to obtain a new one.

Figure 3 represents the UML class diagram defining the holon concept meta-model. In order save place and limit the complexity, in this meta-model, we have not represented the many constraints that apply between classes and that are specified using the OCL language as defined in UML specifications (UML, 2005).

Here is a brief description of this class diagram: The Class *Holon* defines basic attributes for both composite and elementary holons. A *Physical Part* is a reference to the physical part encapsulated in a *holon*. An *Elementary Holon* is defined as a holon with no indication about his lifecycle. For example a product, produced by external manufacturing systems does not give information about the processes needed for its manufacturing. A *Composite Holon* is a holon that has been obtained by either by assembling existing holons, or by disassembling existing holons into new ones.

The *state* class defines the different states that have been observed during the processing phase of the holon. Every manipulation of a holon through a process (Process Instance) implies a change in the state of the processed holon. A *Property* of a *holon* contains information that can not be handled only using its *state*. The *Process instance* refers to the execution of a process on a single holon, this class enables description of the execution of the process with high level of detail (e.g.: elapsed time, start and end of the treatment, used equipment, needed personal). A *Process instance* input is a holon state A *Process* describes an internal process that is performed inside the studied domain. The *Resource* class describes resources needed to perform a process instance. A *resource* can be a material resource, a software resource or a human resource. Each *resource* provides a set of capabilities, and each *process* needs some capabilities to be performed.

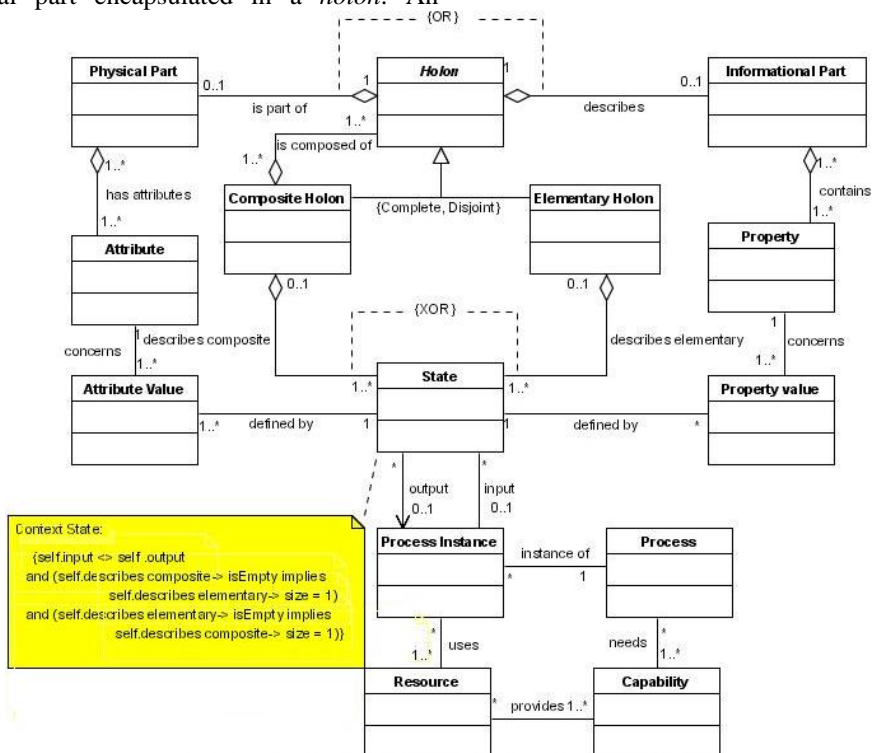

Fig. 3: Class diagram for the Holon model

# 3. HOLONS AND INTEROPERABILITY

The ISO/IEC 23821 Information Technology Vocabulary defines interoperability as "the capability to communicate, execute programs, or transfer data among various functional units in a manner that requires the user to have little or no knowledge of the unique characteristics of those units." The IEEE STD 610.122 standard defines interoperability as "the ability of two or more systems or components to exchange and use information". In this paper, interoperability definition is adapted from the two previous definitions as:

**Definition 1:** Interoperability is the ability to communicate, to cooperate and to exchange models between two or more applications despite differences in the implementation languages, the execution environments, or the models abstraction (Kalfoglou and Schorlemmer , 2004)

Interoperability can be classified into two categories considering the enterprise hierarchy model:
"Horizontal Interoperability" is the interoperability between applications from the same conceptual level in the enterprise. This first category of interoperability aims to synchronise models that were created in different enterprises even those managed by different modelling systems (e.g.: enabling organisational interoperability between two systems used in two different organisations).
"Vertical interoperability" is the interoperability between applications from different enterprise levels. The objective of this category of interoperability is to maintain coherence between information that is handled in two different level of the enterprise (e.g.: ensuring coherence between organisational models of the enterprise and the process models used at shop floor level).

The following introduces the Levels of Conceptual Interoperability Model (LCIM). Similar to the technical approaches, five levels of interoperability are defined (Tolk and Muguira, 2003). The focus lies on the data to be interchanged and the interface documentation, which is available. The layers are defined as follows:

**Level 0 - System Specific Data:** No interoperability between two systems. Data is used within each system in a proprietary way with no sharing. The component (or application) is a black box.

**Level 1 – Documented Data:** Data is documented using a common protocol and is accessible via interfaces. The component is a black box with an interface.

**Level 2 – Aligned Static Data:** Data is documented using a common reference model based on a common ontology, i.e., the meaning of the data is unambiguously described. This is also possible by using metadata standards or by using standard reference models. The component is a black box with a standard interface.

**Level 3 – Aligned Dynamic Data:** The use of the data within the federate/ component is well defined using standard software engineering methods such as UML. This shows the use of data within the otherwise unknown "black box behind the interface," also known as white box.

**Level 4 – Harmonized Data Semantic:** connections between data that are not related concerning the execution code is made obvious by documenting the conceptual model underlying the component.

In order to take into account interoperability requirements during modelling phase in the context of manufacturing systems, we introduce, in this section, the holonic modelling approach for interoperability. Existing interoperability standards and most of existing techniques that enable business process or workflow interoperability are based on a message exchange paradigm (e.g. Wf-XML, BizTalk, FIPA ACL.). These solutions resolve only the particular case of syntactic interoperability (messages vocabulary, messages format, data types, etc). In this section, we show how the holon concept can be used as a means for resolving interoperability issues. First, we will show the use of the holon to handle horizontal interoperability concerns at modelling time. Second, the case of vertical interoperability is studied.

## 3.1 Holon in action for horizontal interoperability

Horizontal interoperability problem occurs when two or several systems or applications from the same level in the enterprise hierarchy (see figure 1) need to exchange information or data in order to perform a common objective. For example, we consider the case of a manufacturing shop-floor where several manufacturing systems need to cooperate in order to achieve a common goal, the release of the final product to be specific. In this section, we show how the use of the holon concept in the modelling phase, enables considering vertical interoperability concerns at modelling time; in the aim to facilitate resolving interoperability problems during engineering phase.

To model manufacturing shop-floor, we use a minimal business process meta-model composed of four Entities:

*Actor*: represents a person or a group of persons that act in someway on processes or in the information system of the enterprise. An a actor can be internal or external to the enterprise

*Process*: is a value chain that provides a good or a service to an internal or external customer.

*Site*: a geographic place where the enterprise is established. Sites can express a special kind of places such as agency, office and factory, or can also express precise geographic places.

*Flow*: is a set of elements (data, information, energy, material ...) that are exchanged between processes

To those entities, we add the notions of *Holon* which *represents* products instances. As we see in section 2.2, a *holon* is described by *properties* and *attributes* that are mandatory for controlling the execution of a process on the *holon*. To manipulate those pieces of information we assume that each process is indeed composed of two interdependent sub-processes: *(i)* An informational process is responsible of manipulating, updating and controlling the information concerning the product (holon), this informational process can be implemented by an application that is performed on the information contained in the product, *(ii)* a physical process that performs all physical transformations on the material of the product. Those two sub-processes are performed in an atomic operation (both are executed or none).

Two types of relationships between a process and a piece of information (property or attribute) have been identified: production and consumption;
- *Production:* we say that a process produces an attribute (or property) when the attribute did not exist before the execution of the process;
- *Consumption:* a process is said to be consumer of an attribute (or property) when it uses the attribute (or property) or updates it.

The specification of relationships between processes and pieces of information during modelling phase enables defining the interfaces of processes at modelling time. The interface of a process defines its inputs and outputs.

Using those interfaces, interoperability of processes using can then be defined as explained in the following:

**Definition 2:** A process P is said interoperable with a system S (composed of processes) iff each input of P is declared as an output of one of his predecessors in S.

The precedence relation between processes is defined as following:

**Definition 3:** The relation of precedence is partial order between processes; we say that a process P1 precedes a process P2 ($P1 <_{Pred} P2$) if it exists a path composed of flows and processes that leads from P1 to P2. In the case cyclic systems, occurrences of execution of processes should be considered; example $P1_i <_{Pred} P2_i$ the i$^{th}$ execution of P1 occurs before the i$^{th}$ execution of P2.

Using the holonic modelling concepts in manufacturing context, enables the considered process interoperability to be concerned at modelling time and not during the engineering phase. This interoperability is a vertical integration of processes, since all process (informational and physical) involved in the studied system are from the same enterprise level, the process control level to be specific. The obtained interoperability is categorised into level 1 of the LCI model (see section 3), it defines interfaces for shop floor process, that are seen as black boxes, since the designer does not know in advance their internal structure and characteristics.

## 3.2 Vertical interoperability with the MDA approach

In this section, we introduce an approach for interoperability based in a model driven architecture (MDA) (Breton and Bézivin, 2001; Mellor, *et al.*, 2004). The main objective of this section is to show how models based on the holon concept defined in section 2.1 could be expressed and transformed into models based on existing data exchange standards and other unified languages.

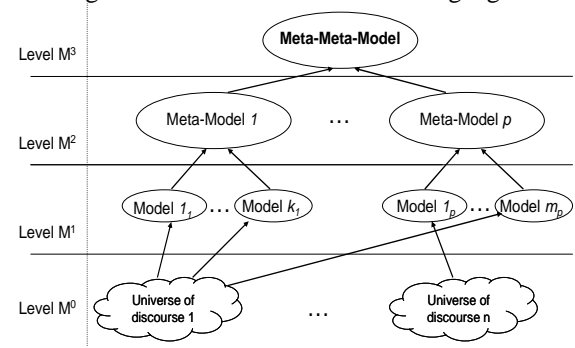

Fig. 4 The four-level ontological approach.

Figure 4 shows the four-level ontological approach levels for modelling that are used in the MDA. As it is explained in (Naumenko and

Wegmann, 2003), the lowest level $M^0$ presents different subjects for modelling, called universe of discourse. The level $M^1$ contains different models of each universe of discourse. The next level $M^2$ presents domain specific meta-models: one meta-model for each of the domains of interest relevant for the $M^1$ models. And finally, $M^3$ level presents a meta-meta-model designed to allow the definition of all the existing in the scope of the meta-models. In this context, applications interoperability may be solved by a top-down approach based on the four levels of the MDA. Indeed the MDA approach for interoperability relies on meta-models mapping to determine, establish and measure interoperability between applications. Several research works have been done in order to resolve meta-models mappings, more generally ontology mappings problems (Kalfoglou and Schorlemmer, 2004).

R Lemesle, in (Lemesle, 1998), explains how models transformation can be resolved by establishing transformation rules between meta-models. Those transformation rules define a mapping that guides model transformations from the instances of the source meta-model to instances of the target meta-model. Those mappings are the bases for applications interoperability. In the MDA approach for applications interoperability, we consider that each application is based on a specific meta-model; Let us consider two applications A and B: A and B are interoperable, if and only if there is a mapping from the meta-model of A ($M_A$) to the meta-model of B ($M_B$) and a mapping form $M_B$ to $M_A$. Those mappings ensure that we can build a model compatible with A from a model used by B (and vice versa).

In Order to use the MDA approach for interoperability in the holonic context, we need to define roles played by the holon in this structure, and to position the holonic modelling approach in terms of models, meta-models and universe of discourse: M2, M1 and M0 (see Fig. 5). In the holonic context, the universe of discourse M0 concerns "The Manufacturing Enterprise Product Universe", to describe this universe of discourse we use holonic models (M1) that are instantiations of the holonic meta-model defining holons and their relationships with other entities in their environment (M2).

Defining interoperability mappings between the holonic meta-model and other meta-models that handle product information enables the holonic meta-model to play the role of a gateway between those meta-models. Indeed, the holonic meta-model can be seen as a reference model for product representation.

In the next section, an implementation of the holonic model and the interoperability mappings is proposed. This implementation relies on a commercial computer assisted software engineering (CASE) tool.

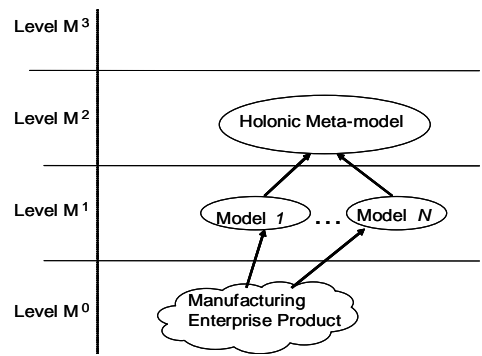

Fig. 5: Holon introduction in the MDA four ontological levels

## 4. IMPLEMENTATION

To experiment the holonic approach defined above in real case we have implemented this approach into a commercial CASE tool named MEGA Suite[6]. MEGA is an enterprise process modelling environment that contains a business process analysis and process modelling and design tools. MEGA has its own meta-model that described all concepts and objects ready to use in MEGA, and all relationships that exist between those concepts. This meta-model can be customized and specialised for specific users needs. MEGA Suite can be used to define, describe and exploit several kinds of diagrams (e.g: Business process Diagrams, UML Diagrams, Workflows). In our contribution, we focus only on business process diagrams; indeed they seem to be the most adequate choice for holon integration. Business Process diagrams in MEGA are based on a meta-model inspired from BPMN[7].. MEGA offers tools that enable customizing the meta-model; we used these tools to embed our own holon meta-model into the existing meta-model of MEGA in order to test the usability of our proposal.

The example presented in Figure 6 shows an example of models that can be designed using the holon modelling concept to represent products in a manufacturing process model. For the sake of simplicity, this example contains only one single process that takes a holon flow as input, and produces a holon flow as output.

---

[6] MEGA Suite, MEGA International, *www.mega.com*
[7] Business Process Modelling Notation, *www.bpmn.org*

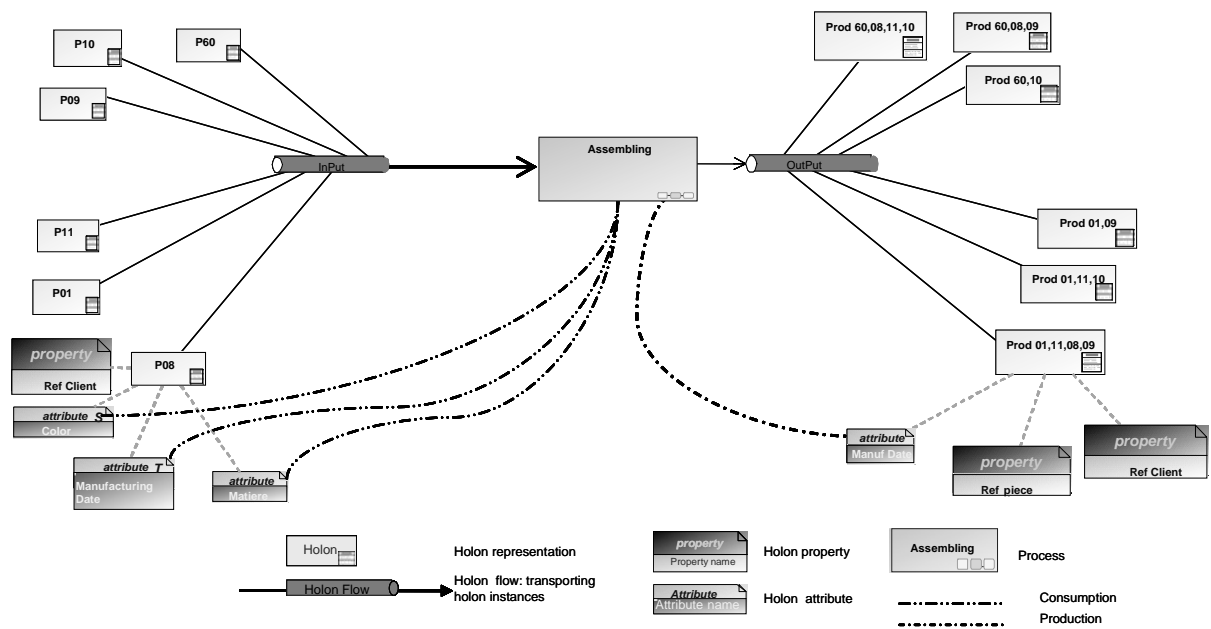

Fig. 6 An example of models containing holons.

In this example, we show using the implementation of the holonic concepts in MEGA, how a process can be connected to information and data concerning holons (inputs or outputs). The holons in this example represent products (final and intermediary).

To experiment the holon models interoperability with other enterprise modelling frameworks using the MDA approach, two examples have been chosen; UEML and B2MML. UEML (Berio, *et al.*, 2003; Panetto, *et al.*, 2004) is the Unified Enterprise Modelling Language, it is used at the organisational level of the enterprise. B2MML (2003) is an implementation of the part 1 of the IEC FDIS 62264 standard (IEC 62264, 2002) developed for interfacing the manufacturing control and execution systems with higher level systems. According to the MDA interoperability approach defined in section 3.2, we now define an example of mappings from the holonic meta-model to The UEML and B2MML meta-models.

Mapping Holon with the Unified Enterprise Modelling Language. The Unified Enterprise Modelling Language (UEML) is the result of the UEML project (UEML, 2003). The UEML is an Interlingua between Enterprise Modelling tools. The meta-model of UEML1.0 (Panetto, *et al.,* 2004) defines the set of most relevant concepts and notions for Enterprise modelling.

Mapping with the B2MML language and the IEC 62264 standard. Business to Manufacturing Mark-up Language (B2MML) is an XML implementation of the IEC 62264 part 1. This standard is composed of six different parts designed for defining the models and interfaces between enterprise activities and control activities. Each model concerns a particular view of the integration problem. Those models show increasing detail level in the manufacturing system.

The detail of those mappings has been published in other papers, for further information see Baïna, *et al* (2005). Vertical interoperability that is established by using those mappings is classified in the Level 2 of the LCI model. (see section 3).

To implement the mappings from the holonic models designed in MEGA and the other formats, we first define an extraction format that expresses data extracted from MEGA holon models in order to reuse it in other tools and frameworks based on other meta-models (UEML, B2MML, etc.). To represent the extracted data, we choose the XML language (XML, 2002); since it is considered as the standard application data exchange language by the W3C. MEGA Suite enables XML files generation in respect to a specific structure. XML structures for UEML (Berio et al., 2003), and B2MML (B2MML, 2003) are used to transform the mappings defined below into XSLT rules that can be applied on the files generated by MEGA in order to restructure them into files that respects the UEML structure or the B2MML structure.

## 5. CONCLUSION

In this paper, we defined an approach for specifying the holon modelling concept, it enables maintaining synchronisation between the physical objects and their informational views in manufacturing environment. Then, we introduced how the holon approach can be used for enterprise interoperability issues. Afterwards, an implementation of our approach in a commercial CASE tool is presented. We also establish a translation mechanism based on meta-model mappings that enables applications using the holonic meta-model to exchange models with other applications based on different meta-models, this mechanism is based on the MDA approach for interoperability.

Ongoing works handle experimentation of the overall approach in an industrial case study, this work is used to verify usability and limits of the approach in real larger scale experiments. Tests are organised into two classes, testing the modelling approach in a real industrial environment and testing the interoperability issues; results are to be published in future papers.

## ACKNOWLEDGEMENTS

This work was funded by the European Commission IST 6th framework programme within the framework of the Network of Excellence INTEROP. The authors would like to thank the entire INTEROP core.

## REFERENCES

B2MML, 2003 The World Batch Forum. Business To Manufacturing Markup Language, version 2.0, 2003.

Baïna.S, H. Panetto and G. Morel, 2005. A holonic approach for application interopearbility in manufacturing systems environment. *In Proc of the 16th IFAC World Congress, Prague, July 4-8, 2005.*

Berio G., et al. 2003 D3.2: *Core constructs, architecture and development strategy*, UEML TN IST – 2001 – 34229, March 2003

Breton E and Bézivin, J (2001) "Model-Driven Process Engineering", 25th Annual International Computer Software and Applications Conference (COMPSAC'01), Chicago, Illinois, Etats-Unis, Octobre 2001.

IEC 62264, 2002. IEC FDIS 62264-1:2002. *Enterprise-control system integration, Part 1. Models and terminology*, IEC, Geneva.

Kalfoglou, Y. and Schorlemmer, M., 2004. Formal Support for Representing and Automating Semantic Interoperability. In *Proceedings of 1st European Semantic Web Symposium (ESWS'04)*, pp. 45-61, Heraklion, Crete, Greece.

Kalfoglou, Y. and Schorlemmer, M. 2003. « Ontology Mapping: The State of The Art. ». The Knowledge Engineering Review, 18: 1-31. 2003. Cambridge University Press

Koestler, A., 1967. *The Ghost in the Machine* Arkana, London.

Lemesle, R., 1998. Transformation Rules Based on Meta-Modelling. *EDOC'98, La Jolla, California, 3-5 November 1998, p. 113-122.*

Mc Farlane, D and Bussmann, S. (2000) Developments in holonic production planning and control, *International Journal of Production Plannig and Control*, Vol. 11, N° 6, pp. 522-536

Morel G., H. Panetto H., M.B. Zaremba and F. Mayer 2003. Manufacturing Enterprise Control and Management System Engineering: paradigms and open issues. IFAC Annual Reviews in Control. 27/2, 199-209, December.

Mellor S.J., Kendall S., Uhl A. and Weise D. 2004. *Model Driven Architecture,* Addison-Wesley Pub Co, March, ISBN: 0201788918.

Naumenko, A., Wegmann, A., 2003. Two Approaches in System Modelling and Their Illustrations with MDA and RM-ODP. In *ICEIS'03, 5th International Conference on Enterprise Information Systems.*

Panetto H., Berio, G., Benali, K., Boudjlida, N. and Petit, M., 2004. A Unified Enterprise Modelling Language for enhanced interoperability of Enterprise Models. *Proceedings of IFAC INCOM 2004 Symposium, April 5th-7th, Bahia, Brazil.*

Panetto H. and Pétin, J.F., 2005. Metamodelling of production systems process models using UML stereotypes, *International Journal of Internet and Enterprise Management*, 3/2, 155-169 - Inderscience Publisher, ISSN: 1476-1300, 2005.

Seidel, D. and Mey, M. 1994, IMS - Holonic Manufacturing Systems: Glossary of Terms, In Seidel D. and Mey M. (eds), IMS - Holonic Manufacturing Systems: Strategies Vol. 1, March, IFW, University of Hannover, Germany.

Tolk, A. and Muguira J. A. (2003). The Levels of Conceptual Interoperability Model. *Simulation Interoperability Workshop Orlando, USA, Sept 2003.*

UML. 2005. Unified Modeling Language, UML2 OCL specification, document ptc/05-06-06, OMG.

Valckenaers, P. (2001), Special issue: Holonic Manufacturing Systems, *Computer In Industry*, 46 (3), pp. 233-331.